\documentclass[a4paper,fleqn]{cas-dc}

\usepackage[numbers]{natbib}
\usepackage{gensymb}
\usepackage{amsmath}

\usepackage{array,booktabs}
\usepackage[font=small]{caption}
\usepackage[flushleft]{threeparttable}
\usepackage{tabularx}
\usepackage{amsmath, amsfonts, mathtools, xfrac, empheq}
\usepackage{multirow}
\usepackage{array,ragged2e}
\usepackage{url}

\usepackage{tikz}
\newcommand{\bola}[2][red,fill=red]{\tikz[baseline=-0.5ex]\draw[#1,radius=#2] (0,0) circle ;}

\def\tsc#1{\csdef{#1}{\textsc{\lowercase{#1}}\xspace}}
\tsc{WGM}
\tsc{QE}
\tsc{EP}
\tsc{PMS}
\tsc{BEC}
\tsc{DE}
\begin{document}

\let\printorcid\relax % Remove ORCID footnote

\shorttitle{MigraR: an open-source R-based application for analysis and quantification of cell migration}
%\shortauthors{CV Radhakrishnan et~al.}

\title [mode = title]{MigraR: an open-source, R-based application for analysis and quantification of cell migration parameters}                      
%%\tnotemark[1,2]

%\tnotetext[1]{This document is the results of the research
%   project funded by the National Science Foundation.}

%\tnotetext[2]{The second title footnote which is a longer text matter
%   to fill through the whole text width and overflow into
%   another line in the footnotes area of the first page.}

%\credit{Conceptualization of this study, Methodology, Software}

%\address[1]{Elsevier B.V., Radarweg 29, 1043 NX Amsterdam, The Netherlands}

\author[1,4,9]{Nirbhaya Shaji}
\author[2,3,9]{Florbela Nunes}
\author[3]{M.Ines Rocha}
\author[2,4]{Elsa Ferreira Gomes}
\author[3]{Helena Castro}

%%\fnmark[2]
%%\ead{cvr3@sayahna.org}
%%\ead[URL]{www.sayahna.org}

%%\credit{Data curation, Writing - Original draft preparation}

\address[1]{FCUP – Faculdade de Ciências da Universidade do Porto, Rua do Campo Alegre s/n, 4169-007 Porto, Portugal.}
\address[2]{ISEP – Instituto Superior de Engenharia do Porto, Politécnico do Porto, Rua Dr. António Bernardino de Almeida 431, 4249-015 Porto, Portugal..}
\address[3]{i3S – Instituto de Investigação e Inovação em Saúde, Universidade do Porto, Rua Alfredo Allen 208, 4200-135 Porto, Portugal.}
\address[4]{INESC-TEC – Instituto de Engenharia de Sistemas e Computadores, Tecnologia e Ciência, Campus da Faculdade de Engenharia da Universidade do Porto, Rua Dr. Roberto Frias, 4200-465 Porto, Portugal.}
\address[9]{These authors contributed equally to this work.}
%* These authors contributed equally to this work.

%\cortext[cor2]{Principal corresponding author}
%\fntext[fn1]{These authors contributed equally to this work.}
%\fntext[fn2]{Another author footnote, this is a very long footnote and
%  it should be a really long footnote. But this footnote is not yet
%  sufficiently long enough to make two lines of footnote text.}
%\nonumnote{This note has no numbers. In this work we demonstrate $a_b$
%  the formation Y\_1 of a new type of polariton on the interface
%  between a cuprous oxide slab and a polystyrene micro-sphere placed
%  on the slab.
%  }

\begin{abstract}
Background and objective: Cell migration is essential for many biological phenomena with direct impact on human health and disease. One conventional approach to study cell migration involves the quantitative analysis of individual cell trajectories recorded by time-lapse video microscopy. Dedicated software tools exist to assist the automated or semi-automated tracking of cells and translate these into coordinate positions along time. However, cell biologists usually bump into the difficulty of plotting and computing these data sets into biologically meaningful figures and metrics.

\noindent Methods: This report describes MigraR, an intuitive graphical user interface executed from the RStudio\textsuperscript{TM} (via the R package Shiny), which greatly simplifies the task of translating coordinate positions of moving cells into measurable parameters of cell migration (velocity, straightness, and direction of movement), as well as of plotting cell trajectories and migration metrics. One innovative function of this interface is that it allows users to refine their data sets by setting limits based on time, velocity and straightness. 

\noindent Results: MigraR was tested on different data to assess its applicability. Intended users of MigraR are cell biologists with no prior knowledge of data analysis, seeking to accelerate the quantification and visualization of cell migration data sets delivered in the format of Excel files by available cell-tracking software.

\noindent Conclusions: Through the graphics it provides, MigraR is an useful tool for the analysis of migration parameters and cellular trajectories. Since its source code is open, it can be subject of refinement by expert users to best suit the needs of other researchers. It is available at GitHub and can be easily reproduced. 
\end{abstract}


\maketitle

\section{Introduction}

Cell migration is an increasingly eminent area of research in biomedical sciences, with implications in fields as diverse as infection, immune response, embryo-genesis, and cancer \cite{10.1002, Bros_2019, Schumacher2019, 1101}. The in-depth study of cells' migratory behavior holds promise to shed new light into these processes and ultimately translate into the development of innovative therapeutic strategies to ameliorate pathological conditions stemming from abnormal cell mobility. One powerful tool to study cell migration is time-lapse video microscopy. By imaging moving cells over several hours, this method captures both the temporal and the spatial dynamics of cell migration. Downstream image processing and cell tracking analyses by dedicated software (reviewed in \cite{Masuzzo2016}) then translate image sequences into numerical values. In their simplest version, cell trajectories are displayed as lists of $x$,$y$ pixel coordinates along time. These coordinates, transformed into standard length measurements ($\mu$m), set the basis to compute biological meaningful quantitative metrics of single-cell migration, namely velocity, straightness and direction of movement, which can be ultimately used to draw informed conclusions about the mobile behavior of a population of cells.

Transforming complex data sets into quantitative and insightful results is not an easy task for most biomedical researchers though. To perform basic tasks, such as graphical display of cell trajectories, plotting of Rose cell-tracking wind diagrams, and data refinement (based, for instances on the time window of data acquisition), cell biologists resort to dedicated \textregistered MATLAB and RStudio\textsuperscript{TM} software.
However, the use of these tools requires an expert knowledge in computer science and mathematics rarely met by health scientists. In the last decade, some more accessible tools have been proposed by the community. A few are free and only some of those are open source (e.g. MotilityLab\footnote{\url{http://2ptrack.net/index.html}}/CelltrackR ~\cite{WORTEL2021100003,Wortel670505}, CellMissy~\cite{CellMissy2013} or Ibidi~\cite{Ibidi}).

The software tool presented in this manuscript was developed to meet the need of a group of biomedical researchers to have access to a free, user-friendly application to make analysis of cell motion. From their collaboration with computer scientists, emerged \textbf{MigraR}.

MigraR is cross-platform, with an intuitive graphical user interface, running from open-source R\textsuperscript{TM} and RStudio\textsuperscript{TM}. With no need for expert knowledge in mathematical data analysis software, MigraR drastically simplifies the plotting and analysis of migration parameters and cell trajectories. Besides quantifying speed and straightness, MigraR also provides specific plots of directions and angles, and, importantly, enables users to filter datasets to improve the quality of their output results. Importantly, MigraR is publicly and freely available on a GitHub account to all interested parties to download and use.  Since its source code is also open, it can be subject of refinement by expert users to best suit the needs of other researchers. 

\section{Material and Methods}

\subsection{Related applications}

MigraR was developed to assist biomedical researchers in the analysis of cell trajectories and migration parameters. MigraR does not perform tracking of cells from time-lapse videos; rather, it processes data sheets of $x$,$y$ coordinates along time that are delivered by upstream image processing and cell tracking software (reviewed in \cite{10.1002}). Launched from RStudio\textsuperscript{TM} using the Shinny extension, MigraR is a fully open source, and user-friendly and interactive platform for uncomplicated analysis of cell migration parameters. Compared with analogous free and open source applications MigraR stands out for (Table \ref{tbl_sw_list}):
\begin{enumerate}[i.]
\itemsep=0pt
\item enabling users to filter their datasets by re-scaling parameters such as time, velocity or straightness;
\item displaying the direction and angle of cell trajectories; 
\item running independently of image analysis software (TrackMate~\cite{BTrackMate}, Ibidi Chemotaxis Manual Tracking~\cite{Ibidi} and ADAPT~\cite{Barry2015}, currently not available, are Fiji/~ImageJ~\cite{Fiji} plugins) and/or of proprietary numerical computing platforms (u-track~\cite{uTrack} runs on \textregistered MATLAB).
\end{enumerate}  

\begin{table*}
\caption{Overview of free and open source solutions similar to MigraR.}
%\begin{threeparttable}[t]
\centering
\begin{tabular}{c c c c c c c c c c c c}
\hline
\fontsize{5}{8}\selectfont 
\textbf{Functionalities} & 
\fontsize{5}{8}\selectfont 
\begin{tabular}[c]{@{}c@{}}CellProfiler \\\cite{McQuin2018} \end{tabular}& 
\fontsize{5}{8}\selectfont 
\begin{tabular}[c]{@{}c@{}}Migrationminer \\\cite{migrationminer} \end{tabular}& 
\fontsize{5}{8}\selectfont 
\begin{tabular}[c]{@{}c@{}}Celltrack \\\cite{CellTRack} \end{tabular}&
\fontsize{5}{8}\selectfont 
\begin{tabular}[c]{@{}c@{}}TrackMate \\\cite{BTrackMate}\end{tabular}&
\fontsize{5}{8}\selectfont 
\begin{tabular}[c]{@{}c@{}}u-track \\\cite{uTrack} \end{tabular}&
\fontsize{5}{8}\selectfont 
\begin{tabular}[c]{@{}c@{}}Ibidi \\Chemotaxis\\\cite{Ibidi}\end{tabular}&
\fontsize{5}{8}\selectfont 
\begin{tabular}[c]{@{}c@{}}CellMissy \\\cite{CellMissy2013}\end{tabular}& 
\fontsize{5}{8}\selectfont 
\begin{tabular}[c]{@{}c@{}}CelltrackR \\ MotilityLab \\\cite{WORTEL2021100003,Wortel670505}\end{tabular}&
\fontsize{5}{8}\selectfont 
\begin{tabular}[c]{@{}c@{}}ADAPT \\\cite{Barry2015}\end{tabular}&
\fontsize{5}{8}\selectfont 
\begin{tabular}[c]{@{}c@{}}shinyHTM \\\cite{shinyHTM}\end{tabular}
\\\hline\hline
\fontsize{5}{8}\selectfont 
\begin{tabular}[l]{@{}l@{}}Dataset refinement\\(by parameters re-scaling) \end{tabular} &  
\fontsize{5}{8}\selectfont 
\bola{1pt} & \bola{1pt} & \bola{1pt} & \bola{1pt} & \bola{1pt} & \bola{1pt} & \bola{1pt} & \bola{1pt} & \bola{1pt} & \bola{1pt} \\\hline
\fontsize{5}{8}\selectfont 
\begin{tabular}[l]{@{}l@{}}Display of direction \\and angle of trajectories\end{tabular} & \bola{1pt} & \bola{1pt} & \bola{1pt} & \bola{1pt} & \bola{1pt} & \bola{1pt} & \bola{1pt} & \bola{1pt} & \bola{1pt} & \bola{1pt} \\\hline
\fontsize{5}{8}\selectfont 
 \begin{tabular}[l]{@{}l@{}}Runs dependently of \\image analysis software\end{tabular} &  &  &
 \fontsize{5}{8}\selectfont 
 \bola{1pt}\tnote{1} &  & \bola{1pt}\tnote{2} & \bola{1pt}\tnote{1} & \bola{1pt}\tnote{3} &  & \bola{1pt}\tnote{1} & \\\hline
\end{tabular}
\begin{tablenotes}\footnotesize
\fontsize{4}{8}\selectfont 
  \item[1]ImageJ plugins; 
  \item[2]Runs on MATLAB;
  \item[3]Runs on Java and mySQL.
  \end{tablenotes}
 % \end{threeparttable}
\label{tbl_sw_list}
\end{table*}

\subsection{Software description, download and execution}
MigraR is a graphical user interface (GUI) software  implemented in R using the Shiny framework. MigraR can be accessible online via the Shinyapps.io server \footnote{\url{https://nirbhaya-shaji.shinyapps.io/migrar/}}. It runs in MS Windows\textsuperscript{TM}, Linux and macOS and requires prior installation of R\textsuperscript{TM} and RStudio\textsuperscript{TM} from website \cite{RStudio}. 
%% EFG AUG
%MigraR time complexity is $O$($z.n$), where $n$ is the data set length (all time steps of all the tracks) and $z<=n$, is the number of different straightness values. 
MigraR time complexity is $O$($n$), where $n$ is the data set length (all time steps of all the tracks). 

\begin{figure}
	\centering
		\includegraphics[scale=0.40] {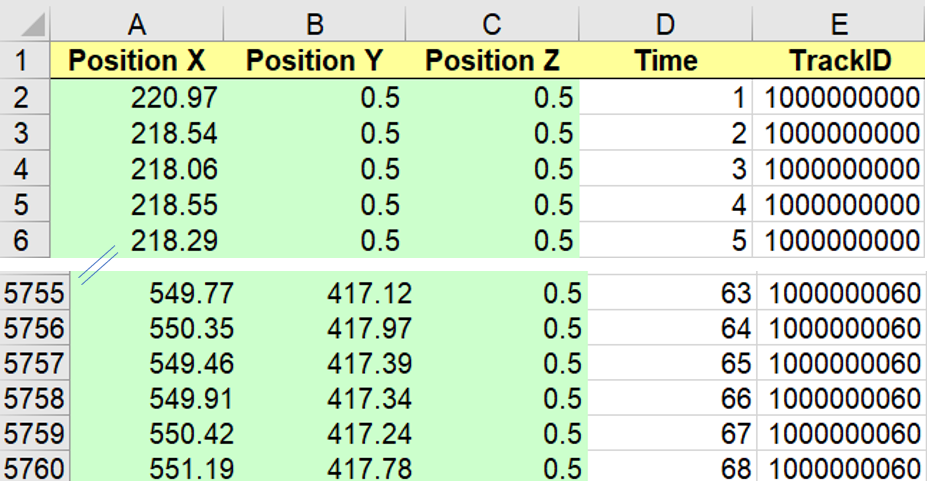}
	\caption{The data sources of MigraR are MSExcel (xlsx, csv) or text files generated by dedicated cell-tracking software; the example provided was deployed by Imaris\textsuperscript{TM}. In these files, data is organized in four columns, as follows: column A) $x$ coordinates; column B) $y$ coordinates; column D) time; column E) TrackID (the cell track identifier, attributed by the cell tracking software).}
	\label{DataSource}
\end{figure}

MigraR sources can be downloaded from GitHub account\footnote{\url{https://github.com/nirbhayashaji/MigraR.git}}. To execute MigraR, users must download the files server.R and ui.R to the working directory of RStudio\textsuperscript{TM} and double click on the server.R file or open it using the file browser form RStudio\textsuperscript{TM}. The file will open in RStudio\textsuperscript{TM}, wherefrom it can be executed by clicking \textit{Run App} \cite{shinyrunapp}. MigraR will open in your default browser. To upload the cell tracking data, users simply have to click on the \textit{Browse} button as seen in (Fig.\ref{DataSource}, top left) listed first on the left panel. Users must be connected to the internet for the initial run so that the required packages can be installed. From there, no active internet connection is necessary for executing the application. The images in the manuscript that corresponds to the MigraR screens are also available in the GitHub for viewing in better clarity. The data used for those images are also available at GitHub. 

\subsection{Data sets}
\begin{figure*}
	\centering
		\includegraphics[scale=0.35] {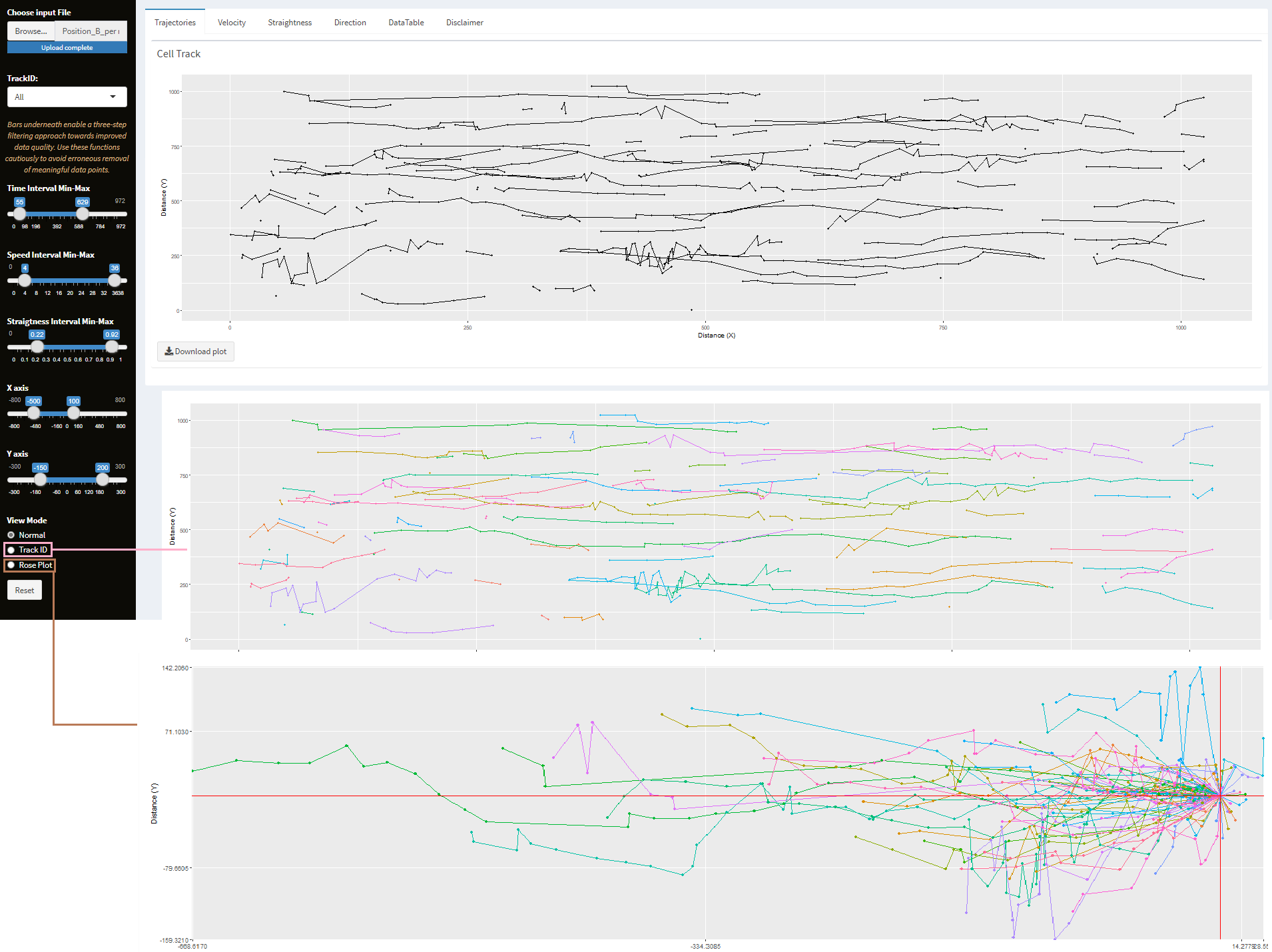}
	\caption{\textit{Trajectories} window from the MigraR interface}
	\label{FigureTraj}
\end{figure*}

The data sources of MigraR are files (txt, csv or xlsx format) generated by dedicated cell-tracking software – an example (provided from Imaris\textsuperscript{TM}) is shown in Fig.\ref{DataSource}. In these files, data must be organized in four columns 
%% added AUG EFG 
, with the fixed titles (as shown in Fig.\ref{DataSource}) and in case of CSV files ',' separated as follows: column A) $x$ coordinates; column B) $y$ coordinates; column D) time; column E) TrackID (i.e., the cell track identifier, attributed by the cell tracking software). Note that decimal separator has to be '.'. The value of TrackID cannot be 0. If it happens MigraR shows a warning message and will ask the user to change the TrackIDs to non zero values. Some cell tracking software might also deliver $z$ coordinates listed in one additional column – column C in the example of Fig.\ref{DataSource}. If the data file does not display $z$ coordinates, users must introduce one dummy (empty or randomly filled) column C. The upload of input files to MigraR is performed by clicking the \textit{Browse} button below the \textit{Choose input file} text as shown in 
Fig.\ref{FigureTraj}.  

\subsection{Computing temporal and spatial coordinates into migration metrics}

MigraR is a digital platform that performs the visual and quantitative inspection of cells' dynamic behavior. It is used after cells trajectories have been recorded by time-lapse microscopy and traced by dedicated cell tracking software. MigraR computes the resulting cells' coordinates along in time, providing users a set of insightful graphical and quantitative outputs.
%%%%%%
MigraR performs a 'cell-based' quantification of cell migration parameters, i.e., it carries out a calculation for each cell separately and, based on each individual result, determines the mean or median for the entire population of cells. This method is in contrast with 'step-based' metrics, for which the mean or median of all separate movement steps are calculated independent of which track it belongs to. Detailed information about advantages and disadvantages of each quantification methods can be found in dedicated literature \cite{Beltman2009, Mokhtari2013}. We should stress that some relevant track analysis metrics, like mean squared displacement (MSD) or autocorrelation plots are not yet supported by MigraR \cite{Masuzzo2016}.
%%%%%%%
One of MigraR functionalities is the graphical display of Rose plots. To draw these graphs, MigraR makes all tracks initiate from the (0,0) coordinate. To do so, the $x$,$y$ coordinates at time zero (the first movie frame) of a given cell track are subtracted to all its $x$,$y$ coordinates along time.
MigraR also calculates the velocity and straightness of cell trajectories, employing the formulas provided in Table \ref{tbl1}, and always taking into account the start and ending points of each cell trajectory. Straightness divides the displacement by the track Length. This gives a number between 0 and 1 (1 means a perfectly straight track).
As we can see in Table\ref{tbl1}, we compute a step-based speed (with one number for each coordinate in the track) and a cell-based approach for the straightness where the total track displacement is divided by the distance travelled in time by the cell (just one number for each track) \cite{Masuzzo2016}.

%%% EFG 5.1 - to complete 
%MigraR calculates step-centric features, computed between two steps of a cell, instantaneous displacements, turning angle and instantaneous speed. Also calculates the cell-centric feature straightness, where the sum and distance travelled in time by the cell straightness index \cite{Masuzzo2016}.

%or other alternatives to straightness 
%- Masuzzo 2015 cited by the authors (Box 2) and doi: 10.1371/journal.pone.0080808

\begin{table*}  
\caption{Equations employed to calculate metrics of migration.}
\centering
\begin{tabular}{ c|l >{\Centering}p{5cm}}
\hline
Parameter & Equation \\
\hline\\ 
Velocity &  \[ v_1 = \frac{\sqrt{(x_{2} - x_{1})^{2} + (y_{2} - y_{1})^{2}}} {t_{2} - t_{1}}, i = 1\]  \\ [4ex]

& \[ v_n = \frac{\sqrt{(x_{n} - x_{n-1})^{2} + (y_{n} - y_{n-1})^{2}}} {t_{n} - t_{n-1}}, i = n\]  \\ [4ex]

&  \[ v_i = \frac{\sqrt{(x_{i+1} - x_{i})^{2} + (y_{i+1} - y_{i})^{2}} + \sqrt{(x_{i} - x_{i-1})^{2} + (y_{i} - y_{i-1})^{2}}} 
{t_{i+1} - t_{i-1}}, 1 < i < n\] \\ [4ex]
%% added AUG EFG 
& \[i=step~time~index,~n=last~time~index~of~the~track\] \\ [2ex]

%new
\hline\\ 
Turning angle & \[\alpha_{i} = tan^{-1}[{(y_{i+1} - y_{i})}/{ (x_{i+1} - x_{i})}] , 1 < i < n\] \\ [2ex]

& \[i=step~time~index,~n=last~time~index~of~the~track\] \\ [1ex]
%new

\hline\\
Straightness & \[S = D/L , D = track~Displacement, L = track~Length\] \\ [3ex]
& \[D=\sqrt{(x_{t_l} - x_{t_f})^{2} + (y_{t_l} - y_{t_f})^{2}}\] \\ [3ex]
& \[L=\sum_{i={t_f}}^{t_l-1}{\sqrt{(x_{i+1} - x_{i})^{2} + (y_{i+1} - y_{i})^{2} }}\] \\[3ex]
& \[t_f = last~time~index~of~track,~ t_l = first~time~index~of~track\] \\ [2ex]
\hline 
\end{tabular}
\label{tbl1}
\end{table*}

\begin{figure}
	\centering
		\includegraphics[scale=1.25] {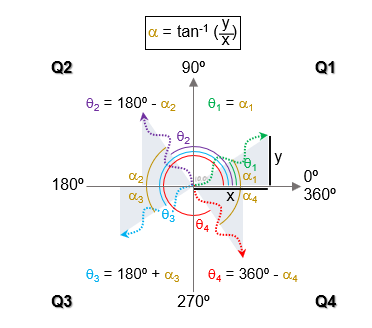}
	\caption{Rationale employed by MigraR to calculate angles of trajectories. The picture represents a $x$,$y$ chart with four fictional cell trajectories (dotted curved arrows), plotted from their standard coordinates (i.e. corrected to start at the 0,0 origin), and designed to end in each one of the four quadrants of the chart (Q1-Q4). To calculate the angle of each trajectory ($\theta$1-$\theta$4), MigraR starts taking the end-point $x$,$y$ coordinates to compute $\theta$1-$\theta$4 angles from the trigonometry function depicted on the top of the chart. The resulting $\theta$1-$\theta$4 angles then are corrected according to the equations depicted in each quadrant (Q1-Q4), to yield the final $\theta$1-$\theta$4 values.}
	\label{angle}
\end{figure}

For calculation of the angle of trajectories ($\theta$), MigraR takes into account $x$,$y$  standard coordinates at the end point of the analysis and uses on the trigonometry equation presented in Fig.\ref{angle}.
%Table \ref{tbl1} (Angle of the trajectory). 
This calculation delivers the value of an angle (angle $\alpha$ in Fig.\ref{angle}). Afterwards, MigraR makes the necessary adjustments according to the quadrant (Q1-Q4) of the $x$,$y$ chart in which cells finalize their trajectories, as follows:
%(Fig.\ref{angle}): 

Q1 (x>0, y>0), $\theta$1 = $\alpha$1;  

Q2 (x<0, y>0), $\theta$2 = 180$\degree$ - $\alpha$2; 

Q3 (x<0, y<0), $\theta$3 = 180$\degree$ + $\alpha$3;  

Q4 (x>0, y<0), $\theta$4 = 360$\degree$ - $\alpha$4.

Angles of trajectories are displayed by MigraR in the form of histograms. In its current version, MigraR accommodate a band width adjustment tool for data refinement based on the number of data points and their distribution - this will avoid biased interpretations, which are particularly tricky in the case of datasets consisting of few cell tracks. For users looking for more refined methods to detect directionality, we suggest alternative tools \cite{Ariotti5285, Textor2011}.

Finally, to have a visual representation of the direction of movement (left or right), MigraR calculates the cosine of $\theta$, and attributes a color code to the resulting values. The cosine value of 1 (yellow color) indicates that cells move to the right, parallel to the $x$ axis; the cosine value of -1 (red color) indicates that cells move to the left, parallel to the $x$ axis; the cosine value of zero (orange color) indicates that cells move up or down. Similarly using the sin value, the vertical movement is represented using colors blue and green for up and down movements respectively.

\subsection{Experimental details for the illustrative data sets used for this report}
The illustrative data set used in this manuscript was obtained by time-lapse microscopy recording of migrating cells, following previously described methodology \cite{Saez2018}. Briefly, murine bone marrow-derived dendritic cells (BMDCs) \cite{AMAN201020}, pulsed for 3 hours with $10\mu g/ml$ LPS were embedded in bovine type I collagen and loaded into a silicon chip mounted on a glass-bottom microscopy chamber slide. Upon polymerization of the collagen/cells mix, medium containing the chemokine CCL21 was added to the microscopy chamber slide. Videos were captured at 10 x magnification in an inverted Leica DMI 6000 FFW (LEICA Microsystems) microscope, equipped with an automated stage, and control of atmosphere  (5\% CO2, 37$\degree C$). The same points in multiple wells were imaged every 4 min for approximately 16 hrs. Movies were subsequently processed in Fiji/ImageJ software \cite{Schneider2012}, prior to automated cell tracking in proprietary software Imaris\textsuperscript{TM} (Bitplane).

\section{Results and Discussion}
\subsection{The four (main) windows of MigraR}
MigraR offers an intuitive graphical user interface for plotting and calculating cell migration parameters. MigraR automatically calculates velocity, straightness and direction of movement based on the spatial and temporal coordinates listed on the uploaded data file. The workflow of MigraR is organized in four windows (\textit{Trajectories}, \textit{Velocity}, \textit{Straightness}, and \textit{Direction}), which can be selected from the respective tabs at the top of the large right side panel (Fig.\ref{FigureTraj}). MigraR also have a window named \emph{DataTable} that lets the user see the raw and calculated data that they are using. 

The \textit{Trajectories} window (accessible from the \textit{Trajectories} tab). The Trajectories window depicts the visual projection of single-cell trajectories (Fig.\ref{FigureTraj}). These can be visualized according to three \textit{View Mode} options, available on the left panel: 
\begin{enumerate}[i.]
 \item the \textit{Normal} view mode is the default option, which provides a close-to-real projection of cell trajectories along the $x$,$y$ axes;
  \item in the \textit{Track ID} mode, individual cell trajectories are distinguished based on color;
  \item in the \textit{Rose Plot} option, the starting positions of each cell trajectory are normalized to $x$,$y$ origin (0,0). Scales of $x$,$y$ axes can be adjusted using scroll bars on the bottom-right.
\end{enumerate}

By clicking on the \textit{Download plot} button at the bottom of the charts, users can download the plot as an image file of \textit{.png} type.

\begin{figure*}
	\centering
		\includegraphics[scale=0.35] {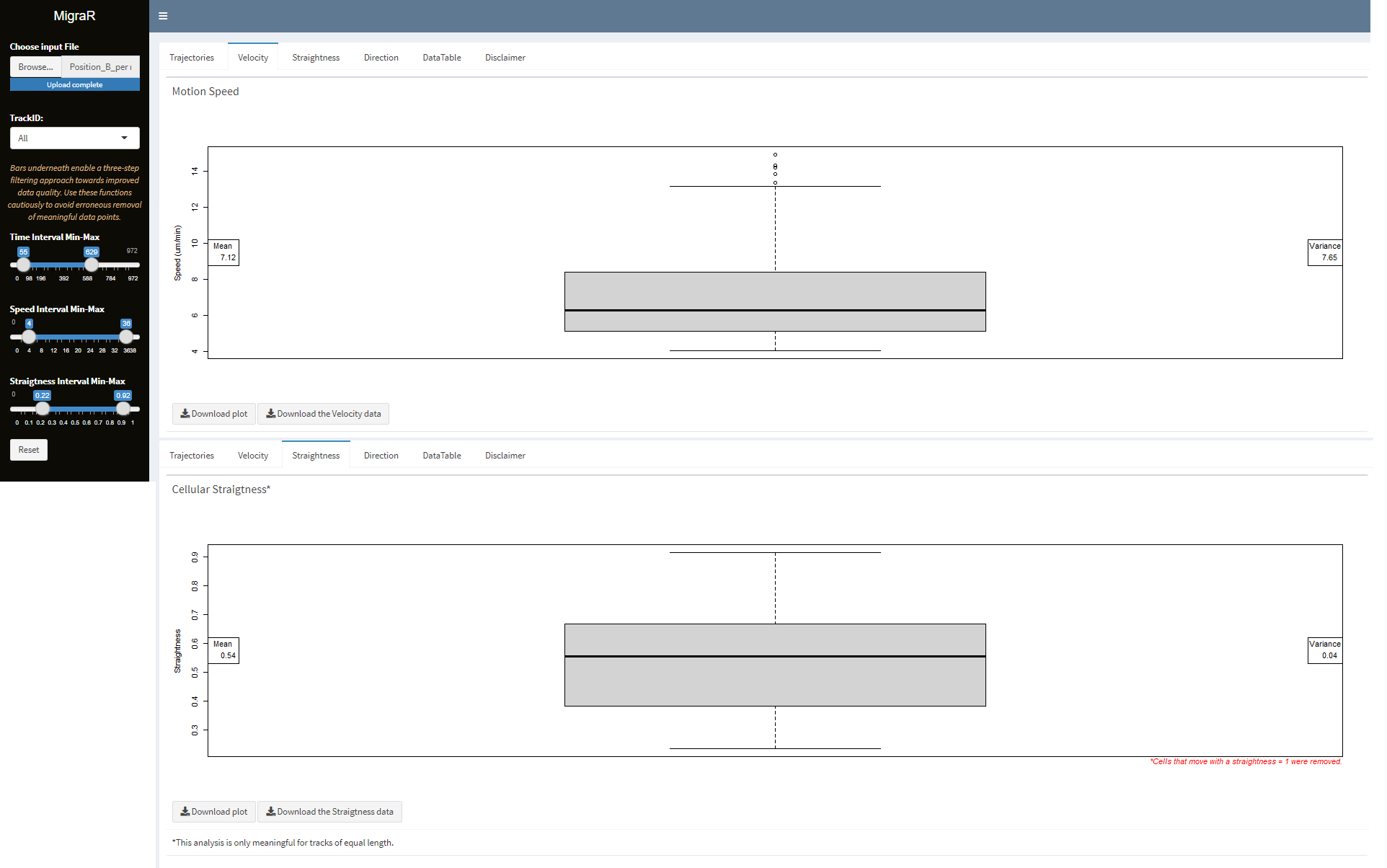}
	\caption{\textit{Velocity} and \textit{Straightness} windows from the MigraR interface.}
	\label{FigureVS}
\end{figure*}

The \textit{Velocity} and \textit{Straightness} windows (Fig.\ref{FigureVS}) display box and whisker charters, plotted from the velocity and straightness values that MigraR computes for set of cell trajectories under analysis. Velocity values refer to the average velocity that each cell assumes along its trajectory. Straightness values refer to the behavior of each cell between the first and the last time point of its trajectory. By pressing the \textit{Download plot} buttons, users can generate and save the plot as \textit{.png} files.
However, the analysis of tracking data is sensitive to biases and artefacts. The user must be careful when using measures like straightness, specially between different datasets \cite{Beltman2009}, \cite{Mokhtari2013}.
%% EFG
In particular, there is a warning message  in the  straightness analysis window that this analysis is meaningful for tracks of equal length.

\textit{Download the Velocity data} and \textit{Download the Straightness data} buttons generate \textit{.csv} files listing the values of velocity and straightness displayed on the corresponding plots. These \textit{.csv} files can be used for subsequent comparative and statistical analyses between different data sets using dedicated software.

\begin{figure*}
	\centering
		\includegraphics[scale=0.35] {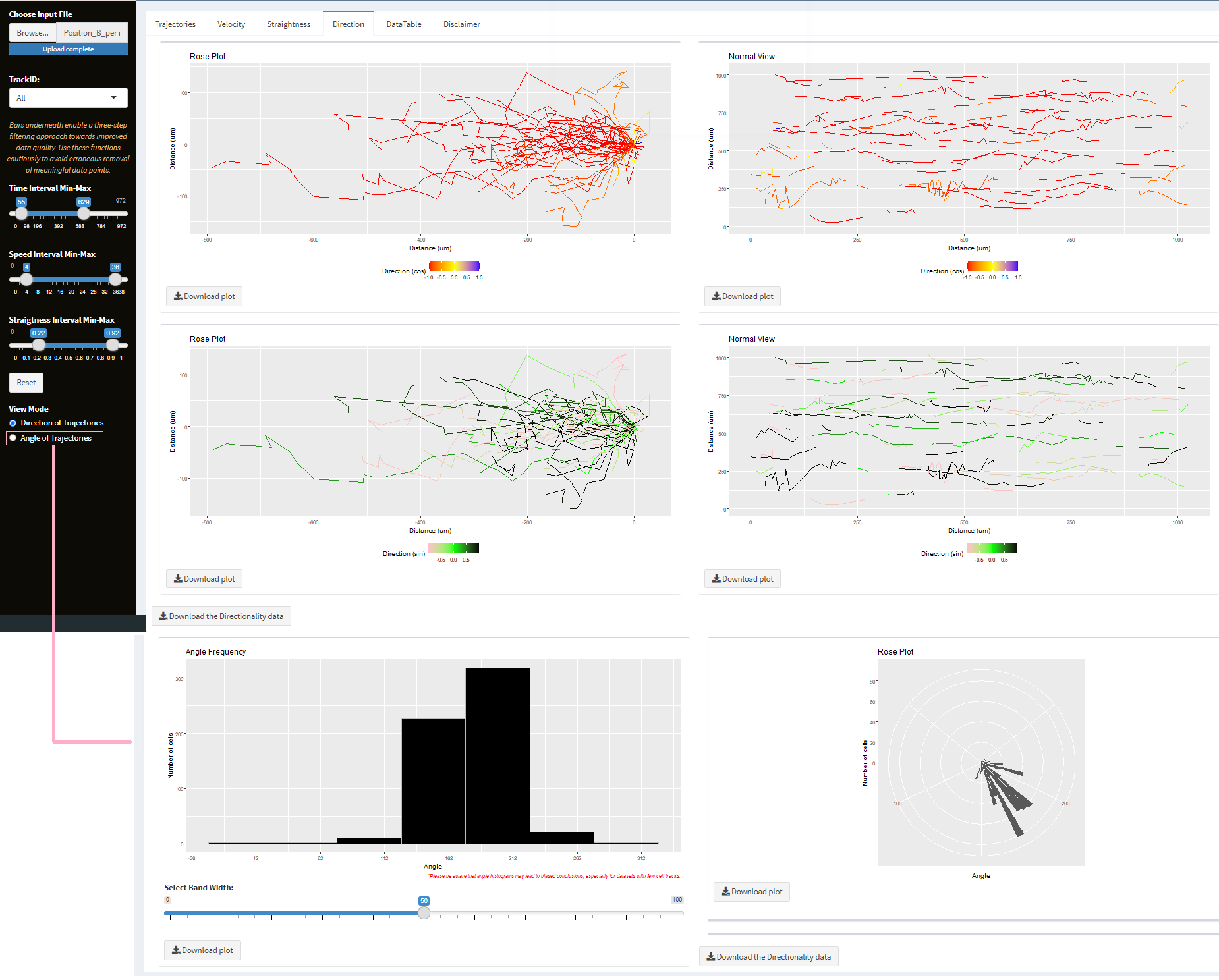}
	\caption{\textit{Directions} window from the MigraR interface.}
	\label{FigureDire}
\end{figure*}

The \textit{Direction} window (Fig.\ref{FigureDire}) is presented with two options, which can be chosen from the radio buttons on the left panel. On the first, \textit{Direction of Trajectories}, users can visualize and download plots where cell trajectories appear in different colors based on the direction of the movement. These cell trajectories plots are available in both their \textit{Normal View} and \textit{Rose Plot} modes. 

In the first pair of plots, cells are colored with a red-to-yellow gradient according to their horizontal directional choices: cells moving to the left are colored red; cells moving to the right are colored yellow. In the second pair of plots, cells are colored with a blue-to-green gradient according to their vertical directional choices: cells moving to the up are colored blue; cells moving to the down are colored green. 

The second option, \textit{Angle of Trajectories}, shows the frequency of cell trajectories moving within the range of  $0-360\degree$ angles, over intervals of 4 degrees. Data are plotted both as column chart (\textit{Angle Frequency}) and \textit{Rose Plot}. These plots, as well as the corresponding data sets are available for download using buttons similar to the ones in the previous windows. 
%EFG
It will be useful if these features were complemented with common track analysis methods not yet included in the tool, displacement analysis, such as mean square displacement (MSD), and angle analysis, direction autocorrelation \cite{Masuzzo2016}.

\subsection{Visualizing Directional and Random movements }

\begin{figure*}
	\centering
		\includegraphics[scale=0.38] {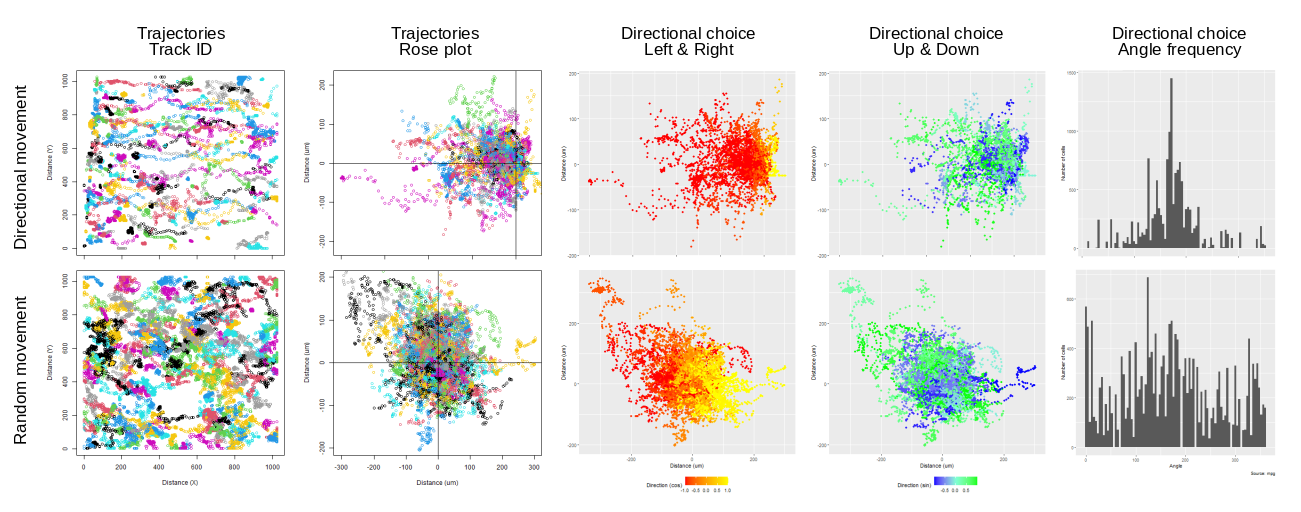}
	\caption{Representative graphical displays produced by MigraR based on two time-lapse microscopy videos: i) 'Directional movement', where cells were stimulated to move in the direction of a chemoattractant placed on their left; ii) 'Random movement', where cells were not exposed to any chemoattractant, hence did not exhibit one prevalent directional choice.}
	\label{FigureFinal}
\end{figure*}

In Fig.\ref{FigureFinal}, we show representative graphical outputs generated by MigraR, starting from two distinct sets of data: one of the datasets (upper panels, ‘Directional movement’) compiles the trajectories of cells exposed to a chemoattractant placed on their left; the other dataset (lower panels, ‘Random movement’) refers to cells that, in the absence of a specific migratory stimulus, moved randomly. Notably, all five graphical displays produced by MigraR translate the different directional choices of cells in each of the experimental conditions under analysis.

\subsection{The adjustments function of MigraR}

All MigraR windows run in parallel with adjustment functions, which can be operated from the left side panel of the graphical interface. Here, users can find three sliders to filter the data sets based on the following parameters: Time, Velocity (Speed), and Straightness as shown in Fig.\ref{FigureDire}. Artifacts are unavoidable in data sets generated by high throughput analysis \cite{Masuzzo2017}. Applying limits or thresholds on these variables is thus desired when cell tracking has been performed fully automatically, albeit it needs to be performed cautiously to avoid elimination of relevant data points. 
With MigraR adjustment functions, users can refine their data sets to exclude outliers resulting from inaccurate tracking [e.g. cells that do not move (most likely dead cells), or to cells that move too fast and/or move with straightness values equal to 1 (corresponding to straight lines, unlikely to occur in nature)]
%With these functions, users can refine their data sets to exclude outliers resulting from inaccurate tracking [e.g. cells that do not move (most likely dead cells), or to cells that move too fast and/or move with straightness values equal to 1 (corresponding to straight lines, unlikely to occur in nature)]. 

Particularly useful is the possibility to filter data sets based on time intervals – with this option, users can look into the dynamics of cell migration at different instants of movie acquisition and chose the most appropriate interval for their analysis. A \textit{Reset} button, at the bottom of this panel, allows users to go back to the original data set. Importantly, these scroll bars work inter-connectedly between windows, i.e. scrolling one of these bars in one window automatically adjusts the same scroll bar in the remaining windows. 

\section{Conclusion}
Cell migration is an increasingly trendy area of research - related publications grew from 1 per year in the early 1970's to >21'000 last year - with potential applications in several branches of health and biological sciences. Powerful commercial tools are available to assist researchers in the complex analysis, visualization and quantification of cell migration. These are nonetheless commercialized at prohibitive prices to groups with poor financial resources. 
%As alternative, researchers can resort to open-source solutions, which are too difficult to use because they require a priori computational and/or mathematical expertise. With MigraR, we expect to democratize the use of cell migration solution by deploying a user-friendly and free solution to all end-users.
As alternative, researchers can resort to open-source solutions, some of which not requiring computational experience (MotilityLab/CelltrackR ~\cite{WORTEL2021100003,Wortel670505}, CellMissy~\cite{CellMissy2013} or Ibidi~\cite{Ibidi}). We expect to democratize the use of cell migration solution with MigraR, a user-friendly, free solution and open-source to all end-users, which has been developed to meet the needs of a group of biomedical researchers.

%%EFG 5.1
As future work, we intend to include some other usual features not yet supported by MigraR.

\textbf{Acknowledgements}.This work is financed by National Funds through the Portuguese funding agency, FCT-~Fundação para a Ciência e a Tecnologia, within project UIDB / 50014 / 2020.

\bibliographystyle{cas-model2-names}

%\bibliography{References}
\bibliography{output.bbl}

\end{document}